\documentclass[prb, showpacs, superscriptaddress, amsmath,amssymb,aps, twocolumn]{revtex4-1}
\pdfoutput=1
\usepackage{txfonts}
\usepackage{amsmath}%
\usepackage{amssymb}%
\usepackage{graphicx}
\usepackage{dcolumn}
\usepackage{bm}
\usepackage{algorithmic}%
\usepackage{color}%
\usepackage{bbm}

\begin{document}

\title{Charge transfer states at the interface of the pentacene monolayer on TiO$_2$ and their influence on the optical spectrum} 

\author{M. P. Ljungberg}
\affiliation{Donostia International Physics Center, Paseo Manuel de Lardizabal, 4. E-20018 Donostia-San Sebasti\'{a}n, Spain}
\affiliation{Department of Physics and Material Sciences Center, Philipps-Universit\"at Marburg, Renthof 5, 35032 Marburg, Germany}
\author{O. V\"ansk\"a}
\affiliation{Department of Physics and Material Sciences Center, Philipps-Universit\"at Marburg, Renthof 5, 35032 Marburg, Germany}
\author{P. Koval}
\affiliation{Donostia International Physics Center, Paseo Manuel de Lardizabal, 4. E-20018 Donostia-San Sebasti\'{a}n, Spain}
\author{S. W. Koch}
\affiliation{Department of Physics and Material Sciences Center, Philipps-Universit\"at Marburg, Renthof 5, 35032 Marburg, Germany}
\author{M. Kira}
\affiliation{Department of Physics and Material Sciences Center, Philipps-Universit\"at Marburg, Renthof 5, 35032 Marburg, Germany}
\author{D. S\'anchez-Portal}
\affiliation{Centro de F\'{\i}sica de Materiales CFM-MPC,
Centro Mixto CSIC-UPV/EHU, Paseo Manuel de Lardizabal 5, E-20018 San Sebasti\'an, Spain}
\affiliation{Donostia International Physics Center, Paseo Manuel de Lardizabal, 4. E-20018 Donostia-San Sebasti\'{a}n, Spain}

\date{\today}

\begin{abstract}
The full monolayer of pentacene adsorbed on rutile TiO$_2$(110) provides an intriguing model to study charge-transfer excitations where the optically excited electrons and holes reside on different sides of the internal interface between the pentacene monolayer and the TiO$_2$ surface. In this work we investigate the electronic properties of this system with density functional theory, and compute its excitonic and optical properties making use of \emph{ab initio} matrix elements.
The pentacene molecules are found to lie flat on the surface, head to tail, and slightly tilted towards the troughs of the oxygen rows of the surface --- in agreement with experiment. Molecular states appear in the band gap of the clean TiO$_2$ surface which enable charge transfer excitations directly from the molecular HOMO to the TiO$_2$ conduction band.
The calculated optical spectrum shows a strong polarization dependence and displays excitonic resonances corresponding to the charge-transfer states. We characterize the computed excitons by their symmetry and location in k-space and use this information to explain the polarization dependence of the optical spectrum.
\end{abstract}

\maketitle

\section{Introduction}

Hybrid semiconducting organic-inorganic systems are currently a subject of great interest~\cite{schlesinger2015efficient,agranovich2016organic, heo2013efficient, verdenhalven2014theory, piersimoni2015charge, brauer2015dynamics, vaynzof2012direct, eyer2015charge}. The leading idea is to combine the benefits of both constituents and devise new applications that outperform their either purely organic or inorganic counterparts. For example, when combining two inorganic materials into a heterostructure it is crucial that the lattice constants between the constituents do not differ too much, and this limits the accessible band gaps for optical applications~\cite{agranovich1998excitons, fluegel2004electronic}. In the organic-inorganic systems, the organic molecules can more flexibly adjust to the underlying lattice of the inorganic substrate, allowing a more flexible tailoring  of electronic band structure~\cite{fluegel2004electronic, wright2012organic}. The band structure of a hybrid system can be further tuned by modifying the properties of the organic material~\cite{schlesinger2015efficient, wright2012organic}. In addition to the band-structure engineering possibilities, these systems can provide new types of semiconductor excitations, like the hybrid Frenkel-Wannier exciton~\cite{agranovich1994hybrid} with a high oscillator strength combined with an enhancement of nonlinear optical response ~\cite{agranovich1998excitons, agranovich2016organic}. Other intriguing excitations include the hybrid charge-transfer exciton~\cite{piersimoni2015charge, brauer2015dynamics, vaynzof2012direct, eyer2015charge} in which a bound electron--hole pair is formed with the electron and hole residing in different sides of the organic-inorganic internal interface of the system.

One particularly successful application of the combination of organic and inorganic semiconductors can be found in photovoltaics ~\cite{wright2012organic}, where dye-sensitized solar cells~\cite{bach1998solid, heo2013efficient, ORegan:1991, Imahori:2009, Hardin:2012, Senadeera:2002} have been found to be promising in the search for efficient, low-cost and environment-friendly devices. The operational principles of solar cells are fundamentally dependent on charge-transfer excitations at the interfaces between the two constituents \cite{eyer2015charge, vaynzof2012direct, brauer2015dynamics, piersimoni2015charge}. Even though charge transfer excitations have been intensively studied during the last decades~\cite{Yu:1995, Zhu:2009, Dou:2013, deibel2010role, benson2007formation, eyer2015charge, vaynzof2012direct, brauer2015dynamics, piersimoni2015charge}, we still lack understanding of many important properties related to the charge-transfer nature of the excited states~\cite{piersimoni2015charge, eyer2015charge, kendrick2012formation}. In order to enhance the performance of photovoltaic devices, it is especially important to characterize the mechanisms involved in the formation~\cite{kendrick2012formation, piersimoni2015charge} and dissociation~\cite{brauer2015dynamics,piersimoni2015charge} of such excitons~\cite{eyer2015charge}. Therefore, it is desirable to search for and study systems where the charge-transfer interface-excitation states are strongly identifiable, e.g., in the optical spectrum. In such systems, we would be able to focus our theoretical and experimental studies on the fundamental character of the charge-transfer states by connecting them to  clearly visible features in the optical spectra.

In many dye-sensitized solar-cells, organic molecules with good light-absorbing properties are placed in contact with porous TiO$_2$~\cite{heo2013efficient, bach1998solid, ORegan:1991}, with the result that an electron is transferred from the molecule to the TiO$_2$ conduction band as a result of the photoabsorption process.  
Pentacene is a $\pi$-conjugated molecule that is of large current interest as electron donor in bulk-heterojuction cells~\cite{Yoo:2004}, often in combination with fullerenes (and functionalized derivatives thereof such as PCBM)~\cite{Yoo:2004}. Also its use in dye-sensitized solar cells has been reported \cite{Senadeera:2002}. When pentacene adsorbs on semiconductor or insulator surfaces, such as SiO$_2$, it is usually almost upright~\cite{Ruiz:2004}, making the $\pi$-orbitals pointing parallel to the surface. However, recently pentacene has been shown to adsorb lying down on the rutile TiO$_2$(110) surface for a coverage up to 2-3 monolayers, as was deduced by STM and X-ray absorption measurements~\cite{Lanzilotto:2011}.  In this configuration, 
 the overlap between the $\pi$-orbitals of the molecule and the substrate states is increased, which could increase the
 interaction strength between the electrons of different constituents, and, consequently, the pentacene monolayer on TiO$_2$ might provide a prototypical system where charge-transfer excitations between the molecule and the surface can take place directly upon optical excitation, enabling in-depth studies of these excitations from the related spectra. Furthermore, experiments~\cite{Lanzilotto:2011} indicate that pentacene molecules form a very well-ordered wetting layer on rutile TiO$_2$(110), providing a well-defined and ordered organic-inorganic interface, making it an ideal candidate for theoretical studies, in contrast to other similar interfaces that show a very complex and disordered structure

In modeling the geometry of molecules on surfaces, density functional theory (DFT) often provides reliable results, with the caveat that van der Waals forces often are  important, especially for weakly adsorbed species. For optical spectra of extended systems, many-body techniques such as the GW together with the Bethe-Salpeter equation \cite{Onida_Reining:2002, bechstedt2015many} or the semiconductor Bloch equations (SBE)\cite{lindberg1988effective, Vaenskae:2015} are commonly used (see appendix \ref{sec:appendix_A} for a summary of the similarities and differences of these approaches). As an input to these methods, we need electronic band energies together with matrix elements for light--matter and (screened) Coulomb interactions. We can obtain these from \emph{ab initio} calculations, from experimentally deduced parameters, or from a combination of both~\cite{Vaenskae:2015}.  

In this work we model the pentacene monolayer on TiO$_2$(110) using density functional theory and compute the optical spectrum from the solution of the Bethe-Salpeter equation for a set of system Hamiltonians. We generate the system Hamiltonians
from \emph{ab initio} data
with an additional and varying parametrization of the dielectric screening, 
allowing us to study the changes in the nature of the dominant optical excitations as the effectiveness of the electronic screening provided by the substrate changes. 
We find that there are optical charge-transfer excitations between the pentacene molecule and the TiO$_2$ surface, demonstrating that the system is promising for studying properties of hybrid charge-transfer excitons. The optical absorption spectra of the charge-transfer states is highly dependent on the polarization of light. To explain the features of spectra, we characterize the symmetries of excitonic solutions and transition dipole elements.

\section{Results and discussion}

\subsection{DFT modeling}
To determine the geometry of the system we optimized it using the \textsc{siesta} DFT code \cite{SIESTA}. We used a slab geometry with the 
1x6 supercell in the surface plane, and five layers of TiO$_2$ in the direction perpendicular to the surface. In total we
have 216 atoms in the unit cell. 20 
\AA  $\,$of vacuum was used in order to minimize the interactions between periodically repeated slabs. 
In order to  include van der Waals contributions to the energy and forces we employed the optB88 functional \cite{Klimes:2010}. As a basis set we used the
DZP basis of numerical atomic orbitals generated using an {\it energy-shift} of 100 meV.
Core electrons were replaced by Troullier-Martins pseudopotentials~\cite{Troullier_Martins:1991}. 
 
\begin{figure}[t!]
 \includegraphics[width=0.5\textwidth]{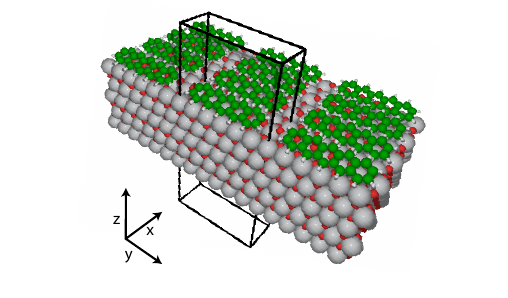}
 \caption{(color online) Geometry of the pentacene/TiO$_2$ slab periodically repeated with the unit cell indicated by a black box. Gray denotes Ti atoms, red O, green C, and white H.}
\label{fig:geom}
\end{figure}

The geometry was optimized with the middle TiO$_2$ trilayer frozen while letting all other atoms, including the molecule, relax. 
The lattice constant in the in-plane direction was fixed to the optimized bulk value obtained with
the same functional and basis set (a=4.60\AA{}, b=2.98 \AA{}), giving a surface 1x6 unit cell of a=6.51 \AA{} 
and b=17.87 \AA{}.  
A model of the used geometry can be seen in Fig. \ref{fig:geom}, where gray denotes Ti atoms, red O, green C, and white H. 
The pentacene molecule is adsorbed over the surface, tilted
inwards to the troughs formed by the Ti-O rows on the TiO$_2$ (110) surface with an angle of 24 
degrees, in agreement with the experimentally deduced 25 degrees\cite{Lanzilotto:2011}. The closest C-Ti distance is 2.76 \AA{}  and the closest C-O distance  (coming from an oxygen row)  is 2.86 \AA.  

%
%

\begin{figure}[t!]
   \includegraphics[width=0.45\textwidth]{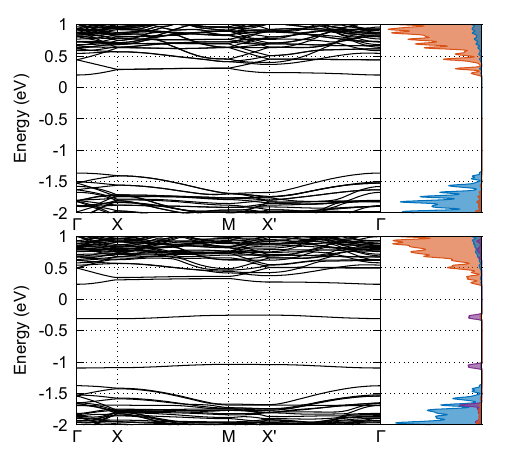}
 \caption{(color online) Band structure (left frames) and projects density of states (right frames) of the pure TiO$_2$ slab (upper frames) and the slab with the pentacene molecule adsorbed (lower frames). In the projected density of states, the colors are  Ti: blue, O: orange, C: purple (the contribution of H is negligible). The fermi levels
of the two structures are shifted to align the highest surface valence band.}
\label{fig:band_structure_one_mol}
\end{figure}

In Fig. \ref{fig:band_structure_one_mol} we show the computed band structure as well as the projected density of states (PDOS) for the different atom species, 
both for the bare surface (upper frame) and the surface with the adsorbed molecules (lower frame).
The high symmetry points in reciprocal space are, in units of the reciprocal lattice vectors, $\Gamma$ = (0, 0, 0), $X$ = (0, $\frac{1}{2}$, 0), $M$ = ($\frac{1}{2}$, $\frac{1}{2}$, 0), and X' = ($\frac{1}{2}$, 0, 0).
As the unit cell is orthogonal, the reciprocal lattice vectors correspond to the same directions in the real lattice shown in Fig. \ref{fig:geom}.
Comparing the surface with and without the molecule, we see that the surface bands are almost unaffected by the adsorption, except for
the appearance of two flat bands in the band gap that are both situated below the Fermi level. Also the lowest conductions band comes down slightly, while still being almost degenerate with another band between the high-symmetry points X and M. 

To assign the bands we take a look at the projected density of states (PDOS). The two gap states below the Fermi level mostly have contributions from the carbon atoms, that is, they belong to the pentacene molecule. The first unoccupied bands, however, belong mostly to the slab and have predominant contribution from Ti 3d orbitals. Indeed, this part of the band structure is very similar to the one coming from the bare surface, without the molecule, so the effect of the molecule on these bands seems to be small. The lowest unoccupied band is seen to be delocalized through the whole slab, so even though it has contributions from the surface atoms it is not a pure surface state. Higher up in energy we see the contributions from the unoccupied pentacene states, centered at around 0.5 eV above the lowest surface conduction band, with the lowest weak feature at 0.25 eV above the same.  The combination of well-separated molecular occupied states and 
and lowest lying unoccupied states belonging to the substrate 
suggests the possibility of optical charge transfer excitations from the molecule to the surface, where the unoccupied molecular states are not expected to contribute much. 

%
%

\subsection{Calculations of excitons and optical spectra}

Linear response optical properties were computed from the Bethe-Salpeter equation \cite{Onida_Reining:2002, Ljungberg:2015} starting from a an effective single-particle system Hamiltonian of the form  $H_{\text{sys}} = H_0 + V_H[G]  +\Sigma[G] $. Here $H_0$ is a zeroth order term independent of the Green's function $G$,   $V_H[G] $ is the Hartree potential and $\Sigma[G]$ is a statically screened exchange potential. 
The linear response of the Green's function can be obtained by solving the eigenvalue problem
%
%
\begin{equation}
H^{BSE}A^{\lambda} = \epsilon^{\lambda} A^{\lambda} \, ,
\label{eq:Matrix_diag}
\end{equation}
where the matrix elements of the effective BSE Hamiltonian $H^{BSE}$ in the one-particle basis (solutions to $H_{\text{sys}}$) are
\begin{equation}
H^{BSE}_{nm\mathbf{k},\, n'm'\mathbf{k'}} = (\epsilon_{m\mathbf{k}}- \epsilon_{n\mathbf{k}})\delta_{nn'}\delta_{mm'}\delta_{\mathbf{kk'}} + (f_{n\mathbf{k}}-f_{m\mathbf{k}}) K_{nm\mathbf{k}, \, n'm'\mathbf{k'}}\, . 
\label{eq:H_eff}
\end{equation}
Here, $\epsilon_{\mathbf{k}}$ are the one-particle solutions to $H_{\text{sys}}$ (a scissor operator can additionally be introduced without loss of generality), $f_{n\mathbf{k}}$ the occupation numbers and $K_{nm\mathbf{k}, \, n'm'\mathbf{k'}}$ the Coulomb kernel for the singlet transition
\begin{equation}
\begin{split}
K_{nm\mathbf{k}, \, n'm'\mathbf{k'}} &= 2\int \psi^*_{n\mathbf{k}}(\mathbf{r}) \psi^*_{m'\mathbf{k'}}(\mathbf{r'}) \, v(\mathbf{r},\mathbf{r'}) \, \psi_{n'\mathbf{k'}}(\mathbf{r'}) \psi_{m\mathbf{k}}(\mathbf{r}) drdr' \\
&- \int \psi^*_{n\mathbf{k}}(\mathbf{r}) \psi^*_{m'\mathbf{k'}}(\mathbf{r'}) \, W(\mathbf{r},\mathbf{r'}) \, \psi_{m\mathbf{k}}(\mathbf{r'}) \psi_{n'\mathbf{k'}}(\mathbf{r}) drdr' \, .
\end{split}
\label{eq:kernel}
\end{equation}
Here $\psi_{n\mathbf{k}}$ are the one-particle wave functions and $v$ and $W$ are the bare and screened Coulomb interactions, respectively. The first term in Eq.~(\ref{eq:kernel}) is the repulsive exchange contribution and the second is the attractive direct term; the one responsible for excitonic binding. 

The optical cross section for Cartesian polarization direction $i$ is related to the longitudinal macroscopic dielectric tensor as $\sigma^i(\omega) = \frac{\omega}{nc} \text{Im} \varepsilon_M^{ii}(\omega)$ where $n$ is the refractive index and $c$ is the speed of light \cite{Onida_Reining:2002, bechstedt2015many}. The macroscopic dielectric tensor is given by $\varepsilon_M^{ii}(\omega) = 1 - \lim_{q_i  \to 0} v(q_i ) \, P^M_{00}(q_i ) $ with $v(\mathbf{q})$ the Coulomb interaction, $P^M_{00}(\mathbf{q})$ is the $\mathbf{G}=\mathbf{G'}=0$ component of the interacting macroscopic polarizability 
\footnote{For a 3-dimensional system the $\mathbf{G}$=0 component should be subtracted for the exchange term in the calculation of $P^M_{00}$, as described in [\onlinecite{Onida_Reining:2002}], [\onlinecite{bechstedt2015many}]}
, and $q_i$ is within the first Brilloiun zone and pointing in the Cartesian direction $i$. In the basis of the eigenstates and eigenvalues of Eq. (\ref{eq:Matrix_diag}) we can obtain the representation
\begin{equation}
\begin{split}
\varepsilon_M^{ii}(\omega) &= 1 -4\pi\sum_{nm\mathbf{k}} D^{i*}_{nm\mathbf{k}} \, \rho_{nm\mathbf{k}}^{i}(\omega) \, ,
\end{split}
\label{eq:epsilon_M}
\end{equation}
using the singlet transition dipole matrix elements 
\begin{equation}
\begin{split}
D^i_{nm\mathbf{k}}  = i \sqrt{2} \frac{\int \psi^*_{n \mathbf{k}}(\mathbf{r})  \hat p_i \psi_{m \mathbf{k}}(\mathbf{r}) dr}{ \epsilon_{m\mathbf{k}}- \epsilon_{n\mathbf{k}}} \, ,
\end{split}
\label{eq:transition_dipoles}
\end{equation}
where $\hat p_i$ is the momentum operator in Cartesian direction $i$ and the $\sqrt{2}$ factor appears for a singlet exciton as described, for example, in Ref [\onlinecite{Ljungberg:2015}]. The density-change matrix $\rho_{nm\mathbf{k}}^{i}(\omega)$ reads
\begin{equation}
\begin{split}
\rho^{i}_{nm\mathbf{k}}(\omega) &= 
\frac{1}{V}\sum_{\lambda, \lambda'} \sum_{n'm'\mathbf{k'}} \frac{ A^{\lambda}_{nm\mathbf{k}} S^{-1}_{\lambda, \lambda'}A^{\lambda' *}_{n'm'\mathbf{k'}} }{\omega - \epsilon^\lambda +i \gamma}  (f_{m'\mathbf{k'}}-f_{n'\mathbf{k'}})D^{i}_{n'm'\mathbf{k'}} \, ,
\end{split}
\label{eq:rho_nmk}
\end{equation}
where the overlap $S_{\lambda, \lambda'}$ between the eigenvectors reflects that the eigenvalue equation in Eq. (\ref{eq:Matrix_diag}) is in general non-Hermitian. Here $V$ is the volume of the supercell. The representation in Eqs. (\ref{eq:epsilon_M}) and (\ref{eq:rho_nmk}) will enable an approximate assignment of the transitions by decomposing the sum in Eq. (\ref{eq:epsilon_M}) in contributions from different $n$, $m$ and $\mathbf{k}$. 
In the Tamm-Dancoff Approximation \cite{Onida_Reining:2002}  (TDA) where particle-hole pairs are assumed to be uncoupled to the hole-particle pairs the solutions of Eq. (\ref{eq:Matrix_diag}) have the meaning of expansion coefficients of an exciton wave function
\begin{equation}
\Psi^{\lambda}(\mathbf{r}, \mathbf{r'}) = \sum_{vc \mathbf{k}}  A^{\lambda}_{vc \mathbf{k}}  \psi_{v\mathbf{k}}(\mathbf{r}) \psi^{*}_{c\mathbf{k}}(\mathbf{r'}) \, ,
\label{eq:exciton_wf_r}
\end{equation}
where indices $v$ and $c$ denote occupied and unoccupied states, respectively. 

Our computational procedure is as follows. First, we rediagonalize the DFT Hamiltonian to obtain wave functions and eigenvalues for all k-points we will use. We assume that the DFT wave functions are close to the solutions to $H_{\text{sys}}$ (an approximation commonly used in GW/BSE calculations) and only modify the band gap using a scissor operator that shifts bulk like bands to have a band gap of 3.5 eV in accordance with experiment
\footnote{Experimentally there exists an uncertainty in the fundamental band gap, for example, it was measured to  3.3 eV in [\onlinecite{Terzuka:1994}]  and to 3.6 in [\onlinecite{Rangan:2010}]}
, giving a shift of 1.63 eV, and a fundamental gap in our system (from the molecular state in the gap to the lowest conduction band) of  2.15 eV. 
In the absence of experimental data regarding the optical band gap in the pentacene/TiO$_2$ interface, our estimation of the band gap must be considered as uncertain, however, within the TDA its exact value will not influence the features of the optical spectrum. 
The matrix elements of the momentum operator in a basis set of local atomic orbitals, which include the corrections due to the use of non-local pseudopotentials, are imported from \textsc{siesta} and are used to compute the matrix elements in Eq. (\ref{eq:transition_dipoles}).
Coulomb integrals are computed by putting the wave functions on a real space grid and solving the Poisson equation for each product state, by going to reciprocal space, and then Fourier transforming back to multiply with the other product state in real space. We use a 2d-truncated Coulomb interaction \cite{Rozzi:2006, Huser:2013} with the cutoff set to half the cell in the normal direction. The divergent Coulomb contributions are numerically integrated \cite{Huser:2013}. We use a plane wave cutoff of 300 eV  for the Coulomb matrix elements (differences in computed spectra with 600 eV were seen to be negligeable). 
 
An important parameter for the qualitative shape of the spectrum is the screening of the direct term in the Coulomb interaction in Eq. (\ref{eq:kernel}). A good model for the electronic part of the screening is the random phase approximation that can in principle be computed \emph{ab initio},  however, the unit cell of our system of study is so large that such a procedure 
would be extremely costly.  
Furthermore, it has been suggested~\cite{bechstedt2015many, Vaenskae:2015} that contributions from phonon modes in bulk TiO$_2$ can be important for the effective dielectric function necessary to properly describe the lowest lying excitons.
In rutile TiO$_2$, there is a large difference between the purely electronic screening 
($\epsilon_{\infty} = $8.43 parallel to the c axis and 6.84, perpendicular to the c axis) 
and the static one where also the nuclei can relax ($\epsilon_0$ =  257 parallel and 111 perpendicular \cite{Madelung:2000}).
 In Ref. [\onlinecite{Vaenskae:2015}] a linear interpolation between  $\epsilon_{\infty}$ and  $\epsilon_0$ was investigated as an effective dielectric constant for the lowest exciton in bulk rutile at the $\Gamma$-point, and there the value of $\epsilon$ =47.8, which is now already geometrically averaged over crystallographic directions, was seen to give a good agreement with experiment.

Also the surface must be taken into account. A simple model of the dielectric behavior of a semiconductor surface is to treat it as a semi-infinite region with a static dielectric constant. In this model the Coulomb interacion will be screened by the arithmetic average of the static dielectric constant with that of vacuum, when $q$ goes to zero \cite{deAbajo:2010}.  In Ref. [\onlinecite{Garcia-Lastra:2009}] an image charge model was used to fit the dielectric constant  for a molecule adsorbed above a TiO$_2$(001) surface with the best fit given by $\epsilon = 2.76$, lower than the average of the bulk $\epsilon_{\infty}$ and the vacuum dielectric constants. In this approach the dependence of the position of the test charge entered as a parameter which makes this model nonlocal, whereas we would like to use an effective local screening.

Given these complications we choose to model the screening by a static dielectric constant with a value set according to some physically motivated model.
 Specifically, we explore three models where the dielectric constant is set to a low ($\epsilon_1$), middle ($\epsilon_2$) and high ($\epsilon_3$) value. For the low value we choose the arithmetic average between the bulk $\epsilon_{\infty}$ (geometrically averaged in the directions) and vacuum, giving $\epsilon_1 = 4.2$, and  for the high value the arithmetic average between the effective bulk dielectric constant seen to give the best agreement for the bulk excitons in Ref. [\onlinecite{Vaenskae:2015}], and vacuum, giving $\epsilon_3 = 24.4$. The middle value of $\epsilon_2 = 7.2$ is chosen to give the average binding energy of the high and the low dielectric constants, assuming the binding energy to be $\propto 1/\epsilon$ as is the case for a two-state system using Eq. (\ref{eq:H_eff}).

 \subsection{Optical spectra}
\begin{figure}[t!]
\includegraphics[width=0.5\textwidth]{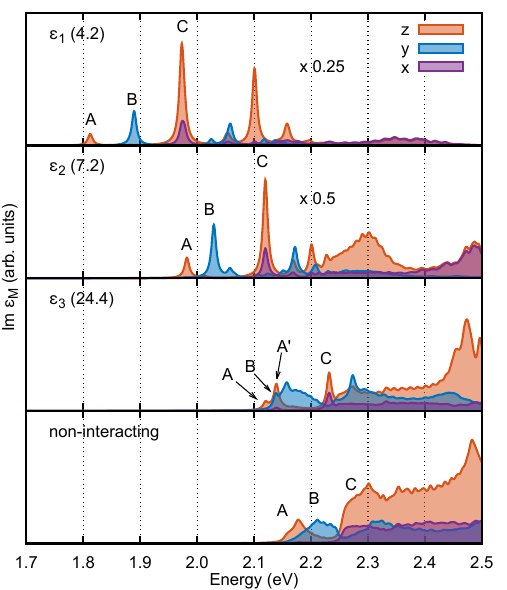}
 \caption{(color online) Calculated macroscopic dielectric function for the pentacene/TiO$_2$ interface (proportional to the optical spectrum) for different polarization directions using various values of the screening parameter $\epsilon$. From the top down we have $\epsilon_1$=4.2, $\epsilon_2$=7.2, $\epsilon_3$=24.4, and the noninteracting result. The different polarization directions are (color online) orange: $z$-polarization, blue: $y$-polarization, purple: $x$-polarization. Also the three low peaks are denoted with A, B and C. For $\epsilon_3$ we also show the extra A' resonance; for clarity its location with respect to A and B are shown with arrows. In all spectra a Lorentzian broadening with a FWHM of 0.01 eV was used.}
\label{fig:spectrum_one_mol_multi}
\end{figure}

In Fig. \ref{fig:spectrum_one_mol_multi} we show the imaginary part of the macroscopic dielectric function (for simplicity denoted spectrum in the following) obtained with the three model dielectric functions, together with the non-interacting spectrum. 

%
%
In all cases, we used one occupied band and five unoccupied bands with a 55x19x1 Monkhorst-Pack k-mesh spanning the whole Brillouin zone (BZ). The Tamm-Dancoff approximation was seen to give visually identical spectra to the full solution of Eq. (\ref{eq:Matrix_diag}) and to obtain an easier interpretation of the solutions we choose to discuss the TDA results throughout this publication. It is clear from Fig. \ref{fig:spectrum_one_mol_multi}  that the Coulomb interaction is important, giving a high sensitivity on the screening model used.
 
While the quantitative spectra vary drastically when going from the noninteracting up to $\epsilon_1$ case, we can identify some qualitative features that seem to be rather independent on the screening of direct Coulomb interaction. Most importantly, the spectra are highly dependent on the polarization of light, so that in addition to the change in absorption intensity between different polarization directions also the energetic location of clearly distinguishable resonances is changed. Most clearly this is seen for the two smaller $\epsilon$ cases. In these spectra, we can clearly separate three excitonic resonances labeled by A, B and C in Fig.~\ref{fig:spectrum_one_mol_multi}. The lowest energy resonance A is clearly visible only for the $z$ polarized light while the resonance B is found similarly connected to $y$ polarization. 
The resonance C yields the leading feature to the spectrum in the cases of $x$ and $z$ polarizations, but practically vanishes for the $y$ polarization.

In the spectra for $\epsilon_3$, we find the same features with a few distinguishing aspects. The A resonance of $\epsilon_1$ and $\epsilon_2$ cases seems to be split to two similarly behaving resonances, which we have labeled by A and A'. 
These resonances are located on both sides of the leading individual resonance B for the y polarization at the low energy region (although in the resolution of our plots AÕ appears at the same position as B, it originates from a separate state).
The stronger $y$-directional absorption slightly above the B resonance originates from multiple weaker resonances that are combined together due to the line broadening we use. The behavior of the C resonance for all interacting cases is highly similar. In the noninteracting system, by definition, we do not have excitonic resonances, but nevertheless even in this case the similar spectral regions corresponding to A, B, and C resonances are detectable, as indicated in the lowest frame of Fig.~\ref{fig:spectrum_one_mol_multi}. 
 
 \subsection{Average hole and electron positions}

\begin{figure}[t!]
\includegraphics[width=0.45\textwidth]{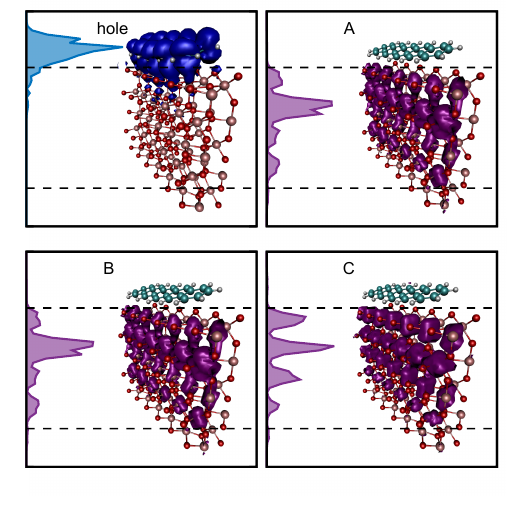}
\caption{(color online) The average electron and hole positions for $\epsilon_1$, as isosurfaces and averaged over the x,y as a function of z. Upper left: the average hole position, for transition A, as a function of z is shown in blue (color online) and its corresponding isosurface in the representation of the unit cell next to it. Upper right: the average electron position for the A transition and its corresponding isosurface are shown in purple. Lower left and right: the same for the B and C transitions. The dashed lines represent the borders of the TiO$_2$ slab.}
\label{fig:average_z_wfs}
\end{figure}

For a given exciton we can compute the averaged density of the electron and the hole from
\begin{equation}
\begin{split}
\rho^{\lambda}_{h}(\mathbf{r})  &=\int |\Psi^{\lambda}(\mathbf{r}, \mathbf{r'})|^2 dr' = \frac{1}{V} \sum_{vv'c \mathbf{k}}  A^{\lambda *}_{v'c\mathbf{k}} A^{\lambda}_{vc\mathbf{k}} u_{v\mathbf{k}}(\mathbf{r}) u^{*}_{v'\mathbf{k}}(\mathbf{r}) \, , \\
\rho^{\lambda}_{e}(\mathbf{r'})  &=\int |\Psi^{\lambda}(\mathbf{r}, \mathbf{r'})|^2 dr = \frac{1}{V} \sum_{vcc' \mathbf{k}}  A^{\lambda *}_{vc'\mathbf{k}} A^{\lambda}_{vc\mathbf{k}} u_{c'\mathbf{k}}(\mathbf{r}') u^{*}_{c\mathbf{k}}(\mathbf{r'}) \, ,
\end{split}
\end{equation}
with $u_{v\mathbf{k}}(\mathbf{r}) $ the cell-periodic part of the one-particle Bloch wave functions used as a basis set for the expansion in Eq.~ (\ref{eq:exciton_wf_r}), the appearance of which reflect the fact that these average densities themselves are cell-periodic. In the limiting case of a single excitation contributing, the particle and hole densities are the square of the single-particle wave functions of the unoccupied and occupied orbital, respectively. In Fig. \ref{fig:average_z_wfs} we show the averaged electron and hole positions over the surface plane direction, as a function of the $z$-coordinate, for $\epsilon_1$ as well as the isosurfaces of the averaged electron and hole wave functions, as represented in the unit cell of the system.  
The averaged hole closely corresponds to the molecular HOMO and is relatively unchanged for different transitions, so we only plot the one for the A transition. For the averaged electron position we show A, B and C transitions. The former two are visually almost identical since they for the most part originate from the same unoccupied band $c_1$.
 The C averaged electron density, however, is quite different. Since the transition mainly comes from the second, $c_2$, band, and although higher in energy, it has part of the density closer to the molecule and consequently has more overlap with the hole state, which partly is an explanation for the higher intensity of this transition. We will later do a more thorough investigation of the transitions strengths by looking at the excitons in k-space.
   
%
%

 \subsection{Polarization dependent absorption and characterization of the excitons}
\begin{figure}[t!]
 \includegraphics[width=0.5\textwidth]{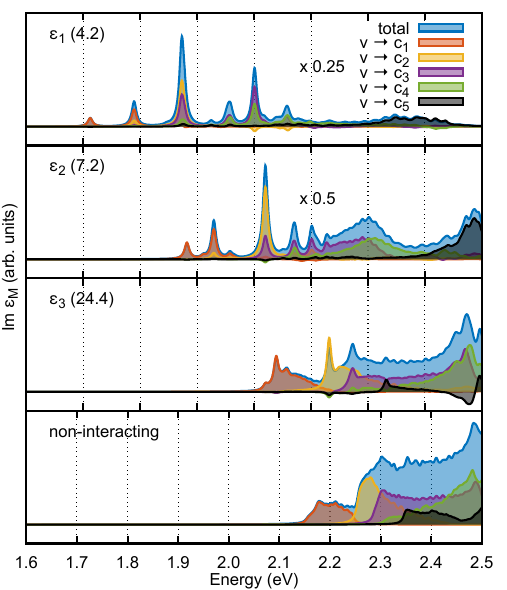}
 \caption{(color online) The approximate band decomposition from Eq. (\ref{eq:rho_nmk}) of the directionally averaged 
  macroscopic dielectric function using various values of the screening parameter $\epsilon$. From the top down we have $\epsilon_1$=4.2, $\epsilon_1$=7.2, $\epsilon_1$=24.4, and the noninteracting result. The colors of the total, and the different conduction band contributions, $c_1$ to $c_5$ are shown in the key of the figure.}
\label{fig:spectrum_decomp}
\end{figure}

To characterize the spectral features seen in Fig.~\ref{fig:spectrum_one_mol_multi}, we first approximately separate the contributions of the different conduction bands, from $c_1$ to $c_5$, to the spectra. Figure~\ref{fig:spectrum_decomp} shows this separation for the polarization-averaged absorption using  a single electron-hole pair in the sum of Eq. (\ref{eq:epsilon_M}). Here the different band contributions sum up to the total spectrum. For the noninteracting case we see that the A and B region in the absorption originate from $c_1$ contributions while the C region comes predominantly from $c_2$. The same holds for $\epsilon_3$, where the A, A' and B excitonic resonances are connected to  the $c_1$-band, while the C resonance originates from $c_2$-band contributions. In the noninteracting case the decomposition is exact, but when the Coulomb interaction is included the density-change matrix in Eq.~(\ref{eq:rho_nmk}) will contain some contributions from other bands due to the eigenvectors $A^{\lambda}_{nm\mathbf{k}}$ that mix the transitions. This also results in some negative spectral features as seen in the interacting cases. Apparently the A resonance seems to originate from the $c_1$-band for all the values $\epsilon$, but this is not entirely true since a spectral feature for a certain band requires a contribution to the exciton eigenvector as well as a non-vanishing transition dipole moment. In fact, looking at the eigenvectors $A^{\lambda}_{v_1c_n\mathbf{k}}$, we see that
when the magnitude of $\epsilon$ is decreased, A and especially B resonances start to have a contribution from band $c_2$, and the $c_3$-band start to have a stronger influence on the resonance C; for $\epsilon_1$ the $c_2$ and $c_3$ contributions are almost equal.

%
%

\begin{figure}[t!]
  \includegraphics[width=0.5\textwidth]{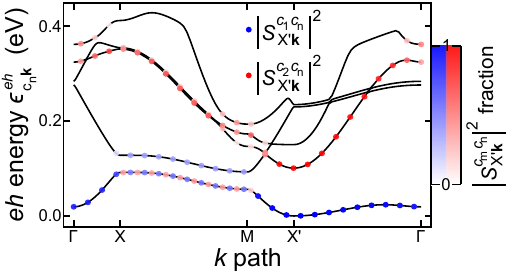}
 \caption{(color online) Band energies relative to the valence band, and squared overlaps of the periodic part of the Bloch wave functions to the one at X'. Blue dots represent the squared overlap to the $c_1$ state and red dots to the $c_2$ state. The strength of the color in the dots represent the magnitude of the squared overlap}
\label{fig:band_mixing}
\end{figure}

To further interpret the polarization dependence and contributions of the spectra in Fig.~\ref{fig:spectrum_one_mol_multi}, we study the dipole, $D^i_{c_nv_1\mathbf{k}}$, and exciton eigenvectors, $A^{\lambda}_{v_1c_n\mathbf{k}}$, as a function of $\mathbf{k}$. These quantities are connected to the relative symmetries of electronic bands and the symmetries of excitons through the periodic part of the electronic Bloch wave functions $u_{n\mathbf{k}}$; $D^i_{c_nv_1\mathbf{k}}$ directly and $A^{\lambda}_{v_1c_n\mathbf{k}}$  via the Coulomb kernel $K_{nm\mathbf{k}, \, nm'\mathbf{k'}}$ . When considering matrix elements across the whole BZ, we face a band-mixing problem where we cannot expect the symmetries of $u_{c_n\mathbf{k}}$ to follow the energetic  ordering of the $c_n$-bands. We illustrate this problem in Fig.~\ref{fig:band_mixing}, by plotting the absolute squares of the overlaps $S^{nm}_{\mathbf{k}'\mathbf{k}}\equiv \langle u_{n\mathbf{k}'}|u_{m\mathbf{k}}\rangle$ with $\mathbf{k'}$ fixed, on top of the electron--hole energies $\epsilon^{eh}_{c_n\mathbf{k}} \equiv \epsilon_{c_n\mathbf{k}}-\epsilon_{v_1\mathbf{k}}-E_g$ for the bands we include in excitonic calculations. Here, $E_g$ is the direct band gap of our system that is located at X'.

%
%

In Fig.~\ref{fig:band_mixing}, we follow $|S^{c_1c_n}_{\text{X'}\mathbf{k}}|^2$ ($|S^{c_2c_n}_{\text{X'}\mathbf{k}}|^2$) by bluish (reddish) dots, which tell how big fraction of functions $u_{c_n\mathbf{k}}$ originate from the function $u_{c_1\text{X'}}$ ($u_{c_2\text{X'}}$), evaluated at $\mathbf{k'}=\text{X'}$. Along the high symmetry paths, the figure shows that the $|S^{c_1c_1}_{\mathbf{k}\mathbf{k}'}|^2$ fraction of the $c_1$ band, which does not have any band crossings, is rather well preserved close to the unity everywhere else except along the X-M path. The situation with higher $c_n$ bands is more complicated. For example, the energetic ordering of the band that mostly resembles $c_2$ at $\mathbf{k}=\text{X'}$ is changing between bands from $c_2$ up to $c_4$; especially challenging is the region around the M point with several band crossings. The band-mixing results in many nonanalytical features in $D^i_{c_nv_1\mathbf{k}}$ and $A^{\lambda}_{v_1c_n\mathbf{k}}$ for individual $c$ bands, when considered across the BZ,  and this complicates the interpretation in terms of $k$-space symmetries. Nevertheless, a look at $D^i_{c_1v_1\mathbf{k}}$ and $A^{\lambda}_{v_1c_1\mathbf{k}}$ terms for $c_1$ band, which conserves its nature across the BZ very well, already gives a lot of information on the nature of A, B, and A' resonances of Fig.~\ref{fig:spectrum_one_mol_multi}.

In order to do this, we need to take into account that the separate diagonalizations of the DFT electronic wave functions at different $k$-points can introduce an arbitrary phase respect to neighboring $k$-points. A similar problem arises from the freedom of selecting the phase factor for excitonic states in Eq.~(\ref{eq:Matrix_diag}).
These problems can be overcome taking into account 
 the following observations regarding our $D^i_{c_1v_1\mathbf{k}}$ matrix elements: firstly, they fulfill 
$D^i_{c_1v_1\,-\mathbf{k}}=(D^i_{c_1v_1\mathbf{k}})^*$, secondly, the absolute values of their real and imaginary parts show continuous behavior, and thirdly, that the imaginary part is (or approaches) zero at the high symmetry points of the BZ. These observations indicate that instead of an arbitrary phase in the DFT wave functions, we only have to deal with an arbitrary sign, which we can fix by 
demanding that $\text{Re}[D^z_{c_1v_1\mathbf{k}}]$ (the most simply behaving $c_1\leftrightarrow v_1$ dipole element) is positive at the $\Gamma$ point and has a continuous behavior from there to across the BZ. For the exciton eigenvectors we use the condition $\text{Im}[A^{\lambda}_{v_1c_1 0}]=0$. As a result of this sign- and/or phase-fixing procedure, we obtain smooth $D^i_{c_1v_1\mathbf{k}}$ and $A^{\lambda}_{v_1c_1\mathbf{k}}$ that have symmetric real parts and asymmetric imaginary parts. Aside from the exciton expansion coefficients $A^{\lambda}_{vc \mathbf{k}}$ it is useful to define an exciton k-space density $\rho^{\lambda}(\mathbf{k})$ by the decomposition 
\begin{equation}
\int |\Psi^{\lambda}(\mathbf{r}, \mathbf{r'})|^2drdr' = \sum_{\mathbf{k}}  \left ( \sum_{vc} |A^{\lambda}_{vc \mathbf{k}}|^2 \right ) = \sum_{\mathbf{k}} \rho^{\lambda}(\mathbf{k})  \, .
\label{eq:rho_lambda_k}
\end{equation}
$\rho^{\lambda}(\mathbf{k})$ is real and non-negative, and summed over $\mathbf{k}$ it is unity for a normalized $\Psi^{\lambda}(\mathbf{r}, \mathbf{r'})$, in other words it closely corresponds to a density in k-space for a given exciton.

%
%
\begin{figure}[t!]
  \includegraphics[width=0.5\textwidth]{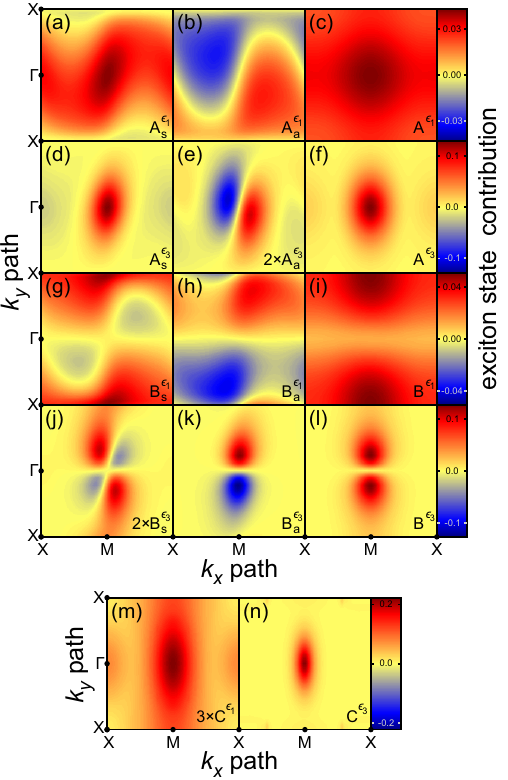}
 \caption{(color online) Exciton eigenvectors and total excitonic $k$-space densities in the Brillouin zone. The $c_1$-band (a) symmetric and (b) antisymmetric contributions for the 
for the $\epsilon_1$ screening and the A excitonic resonance
of Fig.~\ref{fig:spectrum_one_mol_multi}. In (c), the total $k$-space density for the A-resonance $\epsilon_1$ exciton. (d)-(e) show the same results for the $\epsilon_3$ A-resonance, (g)-(i) for the $\epsilon_1$ B-resonance, and (j)-(l) for the $\epsilon_3$ B-resonance. The lowest frames show the total $k$-space densities for (m) $\epsilon_1$ and (n) $\epsilon_3$ C-resonance excitons. The results in (e) and (j) are scaled up by multiplying with a factor of two and the results in frame (m) are scaled up with a factor of three.}
\label{fig:Xs}
\end{figure}
Figs.~\ref{fig:Xs}(a)-(c) show the real and imaginary (symmetric and antisymmetric) contributions of $A^{\lambda;\epsilon_1}_{v_1c_1\mathbf{k}}$ as well as the square root of the exciton k-space density of Eq. (\ref{eq:rho_lambda_k}), for the lowest $\epsilon_1$ exciton. The square root is taken in order to have the same scale for all three plots, and we now explicitly denote the $\epsilon$ used by a superscript. Additionally, we use a simplified notation where $\text{A}^\epsilon_\text{s} \equiv \text{Re}[A^{\lambda;\epsilon}_{v_1 c_1 \mathbf{k}}]$, $\text{A}^\epsilon_\text{a} \equiv \text{Im}[A^{\lambda;\epsilon}_{v_1 c_1 \mathbf{k}}]$, and $\text{A}^\epsilon \equiv \sqrt{\rho^{\lambda;\epsilon}(\mathbf{k})}$, for the $\lambda$ state that corresponds the A resonance of Fig.~\ref{fig:spectrum_one_mol_multi}; similar B and C based notations are used for $\lambda$ exciton states related to the B and C resonances, respectively. In frames (d)-(f) of Fig.~\ref{fig:Xs}, the contributions of the lowest $\epsilon_3$ exciton are shown. 

In general, for all studied $\epsilon$ values 
the lowest excitons show many similar characteristics. They are predominantly (roughly from 60\% to 80\% throughout $c_1$ to $c_5$) coming
 from symmetric $s$-like contributions in k-space centered at X'. The remaining part, from 40\% to 20\% going from low to high $\epsilon$, is given by antisymmetric $p$-like contributions. 
Furthermore, both the symmetric and the antisymmetric parts show a similar tilt with respect to the direction of the reciprocal lattice vectors. 
The A' resonance for $\epsilon_3$ originates from multiple excitonic states that are predominantly symmetric and focused along the X'--$\Gamma$ path. It seems that for smaller $\epsilon$, the X'--$\Gamma$ valley excitons of $\epsilon_3$ are combined in the lowest excitons of $\epsilon_1$ and $\epsilon_2$. This explains the splitting of resonance A of $\epsilon_1$ and $\epsilon_2$ into A and A' in the weaker interaction regime of $\epsilon_3$. 

In frames (g)-(i) and (j)-(k) of Figs.~\ref{fig:Xs}, we show the $c_1$-band symmetric, $c_1$-band antisymmetric and the 
square root of the k-space density
contributions of the excitons that correspond to the B resonance for $\epsilon_1$ and $\epsilon_3$ screening constants. 
For small $\epsilon$ values this state is predominantly (60\%) symmetric and resembles an $s$-like state at M point while having a sizable (40\%) contribution from a $p$-like state located at X' (or at M). The nature of the state drastically changes when the magnitude of $\epsilon$ is increased, so that for $\epsilon_3$ the symmetric fraction is only 20\% and resembles a $d$-like state at X', as seen from Fig.~\ref{fig:Xs}(j). At the same time, in Fig.~\ref{fig:Xs}(k), its $p$-like features centered at X' are strongly increased. 

Figs.~\ref{fig:Xs}(m) and (n) show the  
square root of the k-space density
of $\epsilon_1$ and $\epsilon_3$ excitons related to the C resonances of Fig.~\ref{fig:spectrum_one_mol_multi}. By showing only $\text{C}^{\epsilon}$, we avoid complications from the band-mixings. However, a more detailed study of  
$A^{\lambda;\epsilon}_{v_1c_n\mathbf{k}}$ reveals that the C-resonances excitons are highly symmetric (more or around 80\%) and predominantly $s$-like states centered at X'. The major contribution to these states comes from the $c_n$ bands that resemble $c_2$ at the X' point, and that  
follow the reddish track of dots in Fig.~\ref{fig:band_mixing} along the high symmetry axes.

In other words, in a simple picture, the excitons behind A resonances are \emph{predominantly} $c_1$-band $s$-like excitons located at the X' point, the excitons related to the C resonance are \emph{predominantly} $s$-like $c_2$-band states at X', these two regardless of $\epsilon$, while the exciton state of B resonance changes it nature from a \emph{predominantly} $c_1$-band $s$-like state at M-point to $c_1$ $p$-like state at X' when the magnitude of $\epsilon$ is increased.

In the final stage to interpret the polarization dependency of the spectra in Fig.~\ref{fig:spectrum_one_mol_multi}, we study the combined symmetries of $A^{\lambda;\epsilon}_{v_1c_n\mathbf{k}}$ and $D^i_{c_nv_1\mathbf{k}}$. In here, it is important to note the advantages offered by the symmetric--antisymmetric separation we have done for the dipoles and exciton wave functions. In the oscillator strength sums $\sum_{nm\mathbf{k}} A^{\lambda *}_{nm\mathbf{k}} D^{i}_{nm\mathbf{k}}$ that define our spectra, only the symmetric-symmetric and antisymmetric-antisymmetric parts of products between $A^{\lambda;\epsilon}_{v_1c_n\mathbf{k}}$ and $D^i_{c_nv_1\mathbf{k}}$ contribute. Thus, it is sufficient for us to compare the symmetries and magnitudes of only either symmetric or antisymmetric contributions of $A^{\lambda;\epsilon}_{v_1c_n\mathbf{k}}$ and $D^i_{c_nv_1\mathbf{k}}$ simultaneously. 

%
%
\begin{figure}[t!]
 \includegraphics[width=0.5\textwidth]{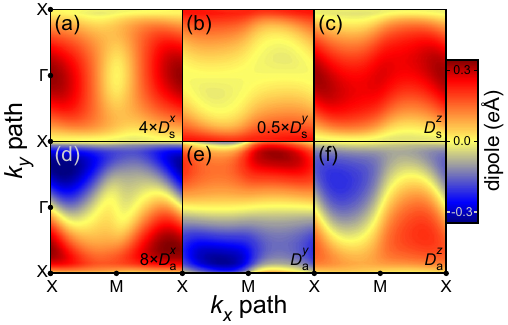}
 \caption{(color online) Transition dipole elements between $v_1$ and $c_1$ bands for (a, d) $x$-, (b, e) $y$-, and (c, f) $z$- polarizations. The dipoles are separated in their symmetric (a)-(c) and antisymmetric (d)-(f) contributions. Dipoles in (a) and (d) are scaled up by multiplying with a factors of four and eight, respectively. Dipoles in (b) are downscaled by multiplying with 1/2.}
\label{fig:DMEs}
\end{figure}

Figure~\ref{fig:DMEs} shows the dipole matrix elements for $v_1$ to $c_1$ transitions. Here, we use notation $D^i_\text{s} \equiv \text{Re}[D^i_{c_nv_1\mathbf{k}}]$ and $D^i_\text{a} \equiv \text{Im}[D^i_{c_nv_1\mathbf{k}}]$. By comparing the symmetric parts of the lowest excitons in Figs.~\ref{fig:Xs}(a) and (d) to the $x$- and $y$-directional dipoles in Fig.~\ref{fig:DMEs}(a) and (b), we notice that the dipoles practically vanish at the X' region where the exciton contributions are centered. The antisymmetric parts of the lowest excitons [Figs.~\ref{fig:Xs}(b) and (e)] show same symmetry as the antisymmetric part of $x$-directional dipole in Fig.~\ref{fig:DMEs}(d), however, the magnitude of this particular dipole term is exceptionally small. The symmetries of $\text{Im}[D^y_{c_1v_1\mathbf{k}}]$, shown in Fig.~\ref{fig:DMEs}(e) and $\text{Im}[A^{1;\epsilon}_{v_1c_1\mathbf{k}}]$ do not match. These observations explain why the A resonances are not visible for $x$ and $y$ polarizations. Especially when comparing Figs.~\ref{fig:Xs}(a) to Figs.~\ref{fig:DMEs}(c) and Figs.~\ref{fig:Xs}(b) to Figs.~\ref{fig:DMEs}(f), one notices the striking similarity between the symmetries of the dipoles and the exciton wave function coefficients. The match of these symmetries make the A resonances relatively well visible in spectra for $z$ polarization.

The same comparison as done above for the A resonance can be made for wave function symmetries of the B resonance in Figs.~\ref{fig:Xs}(g), (h), (j), and (k) with respect the dipoles of Fig.~\ref{fig:DMEs}. This time the match is found only for $y$-polarization, explaining why the B resonance is optically active only for this particular polarization direction. The symmetric dipole-matrix-element contributions that best corresponds to the C resonance show many similarities with the $c_1$ band dipoles in Figs.~\ref{fig:DMEs}(a)-(b), with the exception that the dipole contributions for the C resonance are shifted in the BZ so that the position of X' and $\Gamma$ points are exchanged. In addition, the $x$ and $z$ directional dipoles for C are roughly a factor of seven and a factor of two stronger than the $x$- and $z$-directional dipoles related to $c_1$ band. These remarks explain why the C resonance is visible for $x$ and $y$ polarizations, but is optically inactive for the $y$ polarization, and why the C resonances have a higher intensity compared to A resonances in Fig.~\ref{fig:spectrum_one_mol_multi}.

\section{Conclusions}

We have modeled the pentacene overlayer of the rutile TiO$_2$(110) surface by density functional theory and computed excitons and optical spectra by solving the Bethe-Salpeter equation for a set of physically motivated system Hamiltonians. 
Flat bands coming from the pentacene molecule  are found in the band gap of the pristine surface, and the BSE solutions reveal excitations with charge-transfer character from the molecule to the surface. The optical spectrum shows a large sensitivity on the polarization direction which we explain by investigating the exciton solutions and transition dipole matrix elements in k-space, assigning the three dominant transitions to its approximate symmetry and location in the Brillouin zone. Many qualitative features of the excitons are weakly dependent on the dielectric screening chosen in the system Hamiltonian, although the energetic positions of the peaks are more sensitive to the screening. An experimental characterization of the optical spectrum for this system would help to assess its suitability for photovoltaic applications, or other applications where the created charge transfer excitations could be made use of. 
 
\begin{acknowledgements}
This work was financed by the Deutsche Forschungsgemeinschaft (DFG) through the SFB 1083 project. MPL, DSP and PK also acknowledges support from  Spanish MINECO (Grant No. MAT2013-46593-C6-2-P) and PK also acknowledges financial support from Fellows Guipuzcoa program from Gipuzkoako Foru Aldundia through the FEDER funding scheme of the European Union.
\end{acknowledgements}
\newpage

\appendix

\section{The Bethe-Salpeter equation and the Semiconductor Bloch equations}
\label{sec:appendix_A}

In this work we have been using the Bethe-Salpeter equation to compute the optical spectrum. An alternative method is to solve the Semiconductor Bloch equations which would give almost identical results in the linear regime, and here we outline the similarities and differences between the two methods. 
The BSE takes the form of linear response of the one-particle Green's function with respect to an external perturbation while the SBE \cite{lindberg1988effective} relies on the equation of motion for the microscopic polarization \cite{haug2009quantum, SQO} 
$P^{cv}_k(t) = \langle \hat a^{\dagger}_{ck}(t) \hat a_{vk} (t) \rangle$. 
This quantity is equivalent to a particle-hole matrix element of the density matrix 
$\langle vk | \rho(\mathbf{r}, \mathbf{r'} , t) | ck\rangle$ or indeed that of the 
Green's function
$-i \langle vk | G(\mathbf{r}, t ,\mathbf{r'} , t^+) | ck\rangle $
when the time argument $ t^+$ is infinitesimally larger than $t$. In the BSE, the response of this matrix element determines the polarizability  
when the difference of the two time arguments in $G$ does not play any role, that is, when only a frequency-independent screening is considered in the equations. Thus, when using this standard approximation, the two approaches should be equivalent.  
The optical spectrum in the BSE approach is usually computed from the $q  \to 0$ limit of the dielectric function \cite{Onida_Reining:2002, bechstedt2015many},  that is from the longitudinal response, whereas in the SBE the transverse response is instead used  \cite{haug2009quantum, SQO} .
In the limit of vanishing $q$ the longitudinal and transverse responses coincide.
Since the SBE comes from an equation of motion, it can go beyond the linear response regime in several ways; strong fields that substantially change populations can be used, as well as multiple fields. 
In the linear response regime, the equations of motion can be reduced to an effective eigenvalue problem in the frequency domain which turn out to be equivalent to the corresponding effective eigenvalue problem of BSE, when the standard approximations of diagonal quasiparticle states and static screened interaction are employed.  The screening in the BSE comes out naturally from the theory and requires the direct Coulomb interaction to be screened while the exchange Coulomb interaction should be bare, however for bulk systems the $G=0$ component of this interaction must be disregarded \cite{Onida_Reining:2002, bechstedt2015many} in order to solve for the macroscopic dielectric function $\epsilon^M$ that gives the optical spectrum. In the SBE, usually both the exchange and direct terms are screened, using the rationale of a background dielectric constant coming from inactive electrons that don't contribute directly in the optical process. \cite{SQO. Vaenskae:2015} Since the screening in the SBE is considered to be a parameter it can however also be chosen to coincide with the one of the BSE. 
In summary, the Bethe-Salpeter equation and the Semiconductor Bloch equations are two approaches to the same problem that give very similar (or identical, when using the same approximations) results in the linear regime, while the latter has applicability also for nonlinear phenomena, and the choice to use one or the other would depend on the problem at hand. 

\bibliography{main}{}
\end{document}